\documentstyle[pra,aps,graphicx]{revtex}
\begin{document}

\title{Josephson glass and decoupling of flux lattices in layered 
superconductors}
\author{Baruch Horovitz and T. Ruth Goldin} 
\address {Department of Physics, Ben-Gurion 
University of the Negev, Beer-Sheva 84105, Israel} 
\maketitle 
\widetext
\begin{abstract} 
Phase transitions of a flux lattice in
layered superconductors with magnetic field perpendicular to the 
layers and in presence of disorder are studied. We find that the 
Josephson coupling between layers leads to a strongly pinned Josephson 
glass (JG) phase at low temperatures and fields. 
The JG phase undergoes either a decoupling transition with increasing field or 
a depinning transition with increasing temperature. The resulting 
phases undergo further depinning and decoupling transitions, 
respectively, resulting in a phase diagram with a multicritical point 
where four phases meet. The phase diagram accounts for unusual 
data on $Bi_2Sr_2CaCu_2O_8$ such as the "second peak" transition and 
the recently observed depinning transitions.

\end{abstract}

The phase diagram of layered superconductors in a 
magnetic field $B$ perpendicular to the layers is of 
considerable interest in view of recent 
experiments on high temperature superconductors \cite{Kes}. A first order 
transition in $YBa_{2}Cu_{3}O_{7}$ (YBCO) and in $Bi_2Sr_2CaCu_2O_8$ 
(BSCCO) has been interpreted as a melting transition of the flux 
lattice. This first order transition terminates at a multicritical point, 
which for BSCCO \cite{Khaykovich1,Khaykovich2} is at 
$B_0\approx 300-10^3 G$ and $T_0\approx 40-50 K$,
while for YBCO \cite{Deligiannis} it is at $B_0\approx 2-10 T$ and
$T_0\approx 60-80 K$, depending on disorder and oxygen concentration.
The multicritical point also terminates a "second peak" transition 
\cite{Kes,Khaykovich1,Khaykovich2,Deligiannis} which is manifested by a
sharp increase in magnetization; the transition line at $B\approx 
B_0$ and $T<T_0$ is weakly $T$ dependent. Neutron scattering and $\mu$SR data 
\cite{Kes,Cubitt} show that positional correlations of the flux lattice are 
significantly reduced near these phase boundaries, except however,
near the multicritical point where a reentrant behavior is observed
\cite{Forgan}. Recent data on $Nd_{1.85}Ce_{0.15}CuO_{4-\delta}$ (NCCO) has 
also shown a second peak transition; here, however, $B_0$ decreases with 
temperature near the superconducting transition at $T_c\approx 23 K$ 
with no apparent multicritical point.

In a recent remarkable experiment Fuchs et al. \cite{Fuchs} have 
shown that the phase diagram of BSSCO is much more elaborate. They 
show that the spatial distribution of an external 
 current exhibits a transition from bulk pinning to surface pinning of 
 vortices with most of the current flowing at the sample edges. 
This depinning line crosses the multicritical point and its temperature
is almost $B$ independent at $B<B_0$. The depinning 
transition correlates with anomalies in vibrating reed experiments 
\cite {Kopelevich} and in magnetization \cite{Dewhurst}.
 Thus there are four phase transition 
lines which emanate from the multicritical point at $B_0$, $T_0$: The 
first order line, the second peak line and depinning lines for both 
$B<B_0$ and $B>B_0$.

   An extensive theoretical effort has been devoted to understanding the 
field-temperature ($B-T$) phase diagram \cite{Blatter} in presence of 
disorder. In particular, 
it was proposed that at low $T$ and $B$ a Bragg glass is stable 
\cite{Giamarchi,Kierfeld}, exhibiting algebraic decay of 
translational order and divergent Bragg 
peaks\cite{Korshunov}. Melting 
is expected to occur by thermal or disorder induced dislocations, 
as indeed demonstrated for fields parallel to the 
layers \cite{Carpentier,Golub}.

   The flux lattice can undergo a transition which is unique to 
   layered superconductors, i.e.  a decoupling transition 
\cite{Glazman1,Daemen}. In this transition the Josephson coupling 
between layers vanishes while the lattice can be maintained by the 
electro-magnetic (e-m) coupling between layers. 
 A disorder induced decoupling was also proposed 
as a crossover phenomena \cite{Koshelev1}.

   The theory of Daemen et al.  \cite{Daemen}
    employed the method of self consistent 
harmonic approximation (SCHA) to find the decoupling temperature 
$T_{d}(B)$. The SCHA leads to a conceptual difficulty since it 
predicts that the Josephson coupling vanishes for all purposes at 
$T>T_{d}$. Koshelev \cite{Koshelev2} has shown 
that above some critical temperature the Josephson critical 
current vanishes, however,  a finite Josephson coupling is maintained
 and in fact accounts for the experimentally 
observed plasma resonance. Thus the decoupling transition, as found by 
SCHA, needs to be reinterpreted.

   In the present work we consider temperatures below 
the melting temperature  $T_{m}$ of the flux lattice and study 
(i) the decoupling transition 
in a renormalization group (RG) framework and (ii)
 effects of disorder by employing replica 
symmetry breaking (RSB) methods. We find a glass
transition $T_{g}$ such that for $T<T_{g}$ strong pinning is expected. 
The lines $T_{d}$ and $T_{g}$ cross and lead to four distinct phases 
which meet at one point in the $B-T$ phase diagram,
remarkably close to the experimental phase diagram
\cite{Kes,Khaykovich1,Khaykovich2,Deligiannis,Fuchs,Dewhurst}. 

   Consider a flux lattice with an equilibrium position of the $l$-th 
flux line at ${\bf R}_{l}$. The flux line is composed of a sequence 
of singular points, or "pancake" vortices, whose positions 
at the $n$-th layer can fluctuate to ${\bf R}_{l} + {\bf u}_{l}^{n}$.
 Consider the transverse part of
${\bf u}_{l}^{n}$ with the Fourier transform
 $u_{T}({\bf q},k)$, where ${\bf q}, k$ are 
wavevectors parallel and perpendicular to the layers, respectively. 
The elastic energy due to e-m coupling is given by 
\begin {equation}
{\cal H}_{e-m} = \case{1}{2} \sum_{{\bf q},k} (da^{2})^{2}[c_{66}^{0}q^{2} + 
c_{44}^{0}(k)k_{z}^{2}] |u_{T}({\bf q},k)|^{2}
\end{equation}
where the flux line density is $1/a^{2}$, $d$ is the spacing between 
layers, ${\bf q}$ is within the Brillouin zone (of area $(2\pi 
/a)^{2}$), $|k|<\pi /d$ and $k_{z}=(2/d)\sin (kd/2)$. The shear and 
tilt moduli are given (for $a\gg d$) by \cite{Glazman2}
\begin{eqnarray}
c_{66}^{0} & = & \tau/(16da^{2})  \nonumber\\
c_{44}^{0}(k) & = & [\tau/(8da^{2}\lambda _{ab}^{2}k_{z}^{2})] \ln 
(1+a^{2}k_{z}^{2}/4\pi) 
\end{eqnarray}
where $\tau = \phi_{0}^{2}d/(4\pi ^{2}\lambda_{ab}^{2})$ sets the 
energy scale and $\lambda_{ab}$ is the magnetic penetration length 
parallel to the layers; $\tau \approx 10^3 -10^4 K$ for YBCO or BSCCO
parameters \cite{Kes}. Note the strong dispersion of $c_{44}^{0}(k)$ 
which decreases by the large factor $(d/a)^{2}$ when $k$ varies 
from $k\lesssim 1/a$ to $1/a\lesssim k<\pi /d$.

The Josephson phase between the layers $n$ and $n+1$ at position ${\bf 
r}$ in the layer involves contributions from a nonsingular component 
 and from singular vortex terms. The singular 
phase around a pancake vortex at position ${\bf R}_{l} + {\bf u}_{l}^{n}$ 
is $\alpha ({\bf r} - {\bf R}_{l} - {\bf u}_{l}^{n})$ where $\alpha 
({\bf r}) = \arctan (y/x)$ with ${\bf r} = (x,y)$. We assume that all 
vortices belong to the flux lines, i.e. there are no free 
pancake antipancake ($p\overline {p}$) pairs
 which appear as relevant fluctuations only above $T_{m}$. 
The effect of the nonsingular component is a negligible $T/\tau$ 
term in the RG equation \cite{Horovitz1} while expansion of the 
interlayer phase difference $\alpha ({\bf r} - {\bf R}_{l} - 
{\bf u}_{l}^{n}) - \alpha ({\bf r} - {\bf R}_{l} - {\bf u}_{l}^{n+1})$
yields for the Josephson phase $b_{n}({\bf r}) = 
\sum_{l}({\bf u}_{l}^{n+1} - {\bf u}_{l}^{n}){\bf 
\nabla} \alpha ({\bf r} - {\bf R}_{l})$.
The Hamiltonian is then
\begin {equation}
{\cal H} = {\cal H}_{e-m}
-(J/\xi _{0}^{2}) \sum_n \int d^{2}r \, \cos b_{n}({\bf r})
\label{H}\end{equation}
where $J$ is the interlayer Josephson coupling and $\xi_{0}$ is the 
coherence length, serving as a short distance cutoff. 
Since ${\bf \nabla} \alpha \sim 1/r$ decays slowly, even if ${\bf 
u}_{l}^{n}$ are small the contribution of many vortices which move 
in phase ($q\rightarrow 0$) leads to a divergent response of $b_{n}({\bf 
r})$. In Fourier space, the relevant $b({\bf q},k)$ fluctuations involve
$q \lesssim 1/a$ where \mbox{$b({\bf q},k) = 2\pi 
id(e^{ikd}-1)u_{T}({\bf q},k)/q$}, i.e. enhanced $q\rightarrow 0$ 
fluctuations.

Standard RG proceeds \cite{Horovitz1} by integrating high ${\bf 
q}$ components leading to a new cutoff $\xi>\xi_0$ and a $\xi$ 
dependent coupling 
$J(\xi)$. The significant softening of $c_{44}^{0}(k)$ at $k\gtrsim 
1/a$ implies that the $k$ integration is dominated by $k\approx \pi/d$ 
so that the resulting $c_{66}^{0}[q^{4}/k_{z}^{2}]|b({\bf q},k)|^{2}$ term  
from Eq. (1) can be replaced by an upper cutoff on the $q$ integration, 
$q_{u}=2\log ^{1/2}(a/d)/\lambda_{ab}$. To first order in $J/T$ 
 we obtain $J(\xi)\sim (\xi q_u)^{[1/2(1-t)]}$
where $t=T/T_d$ and the decoupling temperature (similar to the SCHA 
result \cite{Daemen}) is
\begin {equation}
T_{d}=\frac {4a^{4}}{d^{2}}(\int \frac{dk}{c_{44}^0(k)})^{-1} 
\approx \frac{\tau a^{2}\log (a/d)}{4\pi \lambda_{ab}^{2}}
\label{Td}\end{equation}
Thus for 
$T>T_{d}$ $J(\xi)$ vanishes on long scales ($\xi \rightarrow 
\infty$). Second order RG results in 
 renormalization of $c_{44}^{0}$ 
and in generation of Josephson coupling between next nearest 
neighbors \cite{Horovitz1}. The second order terms enhance $T_{d}$ by a 
factor $(1-\gamma J/T)^{-1}$ where $\gamma$ is a nonuniversal 
parameter.

   The RG process shows that the decoupling transition is manifested 
only on long scales. Thus the thermal average of the {\em local}
observable $<\cos b_{n}({\bf r})>$  remains finite at $T>T_{d}$, as in 
the $J/T$ expansion \cite{Koshelev2}. This, however, does {\em not} 
imply long range order in $\cos b_{n}({\bf r})$ - the same $J/T$ 
expansion yields a power law decay for $<\cos b_{n}({\bf r})
\cos b_{n}({\bf 0})>$ correlation, as also obtained from RG.

A hallmark feature of two-dimensional superconductivity is the $\ln 
\rho$ dependence of a $p\overline {p}$ interaction 
on their separation $\rho$, leading to a power law I-V relation 
 \cite{Halperin}. To probe this feature in the decoupled phase we 
 consider a high temperature expansion of  Eq.~(\ref{H}) with an 
 added $p\overline {p}$ pair. The direct e-m interaction 
  is \cite{Horovitz1} $\sim \ln \rho$ while the additional effective
free energy to order $J^2$ is 
\begin {equation}
F_{p\overline {p}}(\rho)\sim (J^2/T)\int d^2r
\int_{|{\bf r}-{\bf r}'|>\xi_0} 
d^2r'|{\bf r}-{\bf r}'|^{-4t}[1-\cos (\alpha_0 ({\bf r},
\mbox{\boldmath $\rho$} )-\alpha_0 ({\bf r}',\mbox{\boldmath $\rho$}))]
\label{pp}\end{equation}
where $\alpha_0 ({\bf r},\mbox{\boldmath $\rho$})=
\alpha ({\bf r}-\mbox{\boldmath $\rho$})-
\alpha ({\bf r})$. Eq.~(\ref{pp}) can be shown to be bounded by a 
$\sim \ln^2 \rho$ term, supporting a nonlinear I-V relation 
at  $T>T_{d}$. In contrast, at $T<T_{d}$ the $p\overline {p}$ 
interaction increases as $\sim \rho$ leading to a finite critical 
current. Thus the decoupling transition is manifested by the change 
in correlation function, vanishing of the Josephson critical current
\cite{Koshelev2} and by nonlinear I-V relation.

We proceed now to study effects of disorder. Since $T<T_{m}$ we 
assume first small fluctuations $|{\bf u}_{l}^{n}|\ll a$. Consider a short 
range pinning potential $U_{pin}^{n}({\bf r})$ which couples to the 
vortex shape function $p({\bf r})$ as
$\int d^{2}r \sum_{n,l} U_{pin}^{n}({\bf r})
p({\bf r} - {\bf R}_{l} - {\bf u}_{l}^{n})$  .
Expansion in ${\bf u}_{l}^{n}$ and averaging $U_{pin}^{n}({\bf r})$ 
by the replica method \cite{Mezard} leads to the replicated Hamiltonian,
\begin {eqnarray}
{\cal H}_r/T&=&\case {1}{2}\sum_{{\bf q},k;\alpha 
,\beta}[c(k)q^{2}\delta_{\alpha ,\beta}-s_{0}\frac{q^{2}}{k_{z}^{2}}]
b^{\alpha}({\bf q},k)b^{\beta *}({\bf q},k)   \nonumber \\ 
& & -\frac{J}{T\xi_{0}^{2}}\sum_{n;\alpha}\int d^{2}r\, \cos 
b_{n}^{\alpha}({\bf r}) - \frac{v}{\xi_{0}^{2}}\sum_{n;\alpha\neq \beta}
\int d^{2}r \cos [b_{n}^{\alpha}({\bf r}) - b_{n}^{\beta}({\bf r})]
\label{Hreplica}\end{eqnarray}
where $\alpha ,\beta$ are replica indices, $c(k)=(a^{2}/2\pi 
d)^{2}c_{44}^{0}(k)/T$ and $s_{0}=\bar{U}a^{2}d/(4\pi d^{2}T)^{2}$ with 
$\bar{U}$ an average of the pinning potential. In Eq.~(\ref{Hreplica}) the 
$c_{66}^{0}$ term has been replaced  by a cutoff $q_{u}$ on $q$ 
integrations, as above. The inter-replica Josephson coupling, 
i.e. the $v$ term in Eq.~(\ref{Hreplica}),
 is generated from the J term in second order RG. It is 
 essential to keep the $v$ term from the start since it couples 
 different replica indices and
can lead to distinct physics by RSB \cite{Golub,Horovitz2}.

Note that the more general form of the disorder term is 
\cite{Giamarchi,Kierfeld,Korshunov} $\cos [{\bf Q}\cdot ({\bf u}_{l}^{n,\alpha}
-{\bf u}_{l}^{n,\beta})]$ where ${\bf Q}$ is a reciprocal lattice 
vector; expansion of this cosine leads to the 
$s_{0}$ term in Eq.~(\ref{Hreplica}). The cosine form is essential 
for deriving the 
Bragg glass properties of the flux lattice, i.e. the $1/r$ decay of 
the displacement correlation at distances $r>\ell$. The domain size 
$\ell$, 
over which the flux lattice is well correlated will be of significance 
below.

The RSB method \cite{Mezard} proceeds by employing a variational free energy 
${\cal F}_{var}={\cal F}_{0}+<{\cal H}-{\cal H}_{0}>$ with ${\cal F}_{0}$
 the free energy of
${\cal H}_{0}=\case{1}{2}\sum_{{\bf q},k;\alpha ,\beta}G_{\alpha ,\beta}^{-1}
({\bf q},k) b^{\alpha}({\bf q},k)b^{\beta *}({\bf q},k)$ and
$G_{\alpha ,\beta}({\bf q},k)$ is determined by an extremum condition 
on $F_{var}$. This yields
\begin {mathletters}
\label{extremum}
\begin{eqnarray}
G_{\alpha ,\beta}^{-1}({\bf q},k)&=&[c(k)q^{2}+z]\delta_{\alpha ,\beta}
-s_{0}(q^{2}/k_{z}^{2})-\sigma_{\alpha ,\beta}   \label{extremuma} \\
z &=& (J/2T\xi_{0}^{2}d)\exp [-\case{1}{2}\sum_{{\bf q},k}G_{\alpha ,
\alpha}({\bf q},k)]   \label{extremumb} \\
\sigma_{\alpha ,\beta} &=& (v/\xi_{0}^{2}d)[\exp (-\case{1}{2}
B_{\alpha ,\beta})
-\delta_{\alpha ,\beta}\sum_{\gamma}\exp(-\case{1}{2}B_{\alpha ,\gamma})]
\label{extremumc} 
\end{eqnarray} 
\end{mathletters} 
where $B_{\alpha ,\beta}=2\sum_{{\bf q},k}[G_{\alpha ,\alpha}({\bf q},k)
-G_{\alpha ,\beta}({\bf q},k)]$ and $z$ is a renormalized Josephson 
coupling. The method of RSB \cite{Mezard} 
represents a hierarchy of matrices such as $\sigma_{\alpha ,\beta},\,
B_{\alpha ,\beta}$ in terms of functions $\sigma(u),\, B(u)$, 
respectively, with $0<u<1$. The amount by which the replica symmetry is 
broken is measured by a glass order parameter $\Delta(u)=u\sigma (u)-
\int_{0}^{u}\sigma(v)\,dv$. Using standard methods 
\cite{Mezard,Horovitz2} we find that
 the solution for $\Delta(u)$ is a step function, 
i.e. $\Delta(u)=0$ for $u<2t$ while $\Delta(u)=\Delta_0$ for 
$2t<u<1$, where
\begin {equation}
(z+\Delta_0 )/\Delta_c=[2tv/\xi_{0}^{2}\Delta_c]^{1/(1-2t)}
\label{Delta}\end{equation}
with the cutoff $\Delta_{c}\approx c(\pi /d)q_u^2$.
Thus a solution with $\Delta_0\neq 0$ is possible only if $t<1/2$. 
 To solve for $z$ in Eq.~(\ref{extremumb}) we need the diagonal part,
\begin {equation}
\sum_{{\bf q},k}G_{\alpha ,\alpha}({\bf q},k)=\ln (2etv/z\xi_0^2 d)
+(s_0/8\pi ^2)[I(z)+zI'(z)]
\label{diag}\end{equation}
where
$I(z)=\int d^2q\,dk/\{k_z^2c(k)[c(k)q^2+z]\}$ and $I'(z)=dI(z)/dz$.
Formally $I(z)$ diverges at $k=0$; this divergence can be traced back 
to our assumption that the $\cos [{\bf Q}\cdot ({\bf u}_{l}^{n,\alpha}
-{\bf u}_{l}^{n,\beta})]$ term is expanded into the $s_0$ term in 
Eq.~(\ref{Hreplica}). Retaining this cosine leads to Imry-Ma type domains
of correlated ${\bf u}_l^n$  
whose size perpendicular to the layers is  $\ell _z$. Within a domain the  
${\bf u}_{l}^{n}$ expansion is valid so that $\pi /\ell _z$ serves as a lower 
cutoff in the $k$ integration. More formally, keeping the 
$\cos [{\bf Q}\cdot ({\bf u}_{l}^{n,\alpha}-{\bf u}_{l}^{n,\beta})]$
term replaces $s_0$ in Eq.~(\ref{extremuma}) by a matrix $\sigma_{\alpha 
,\beta}^{(1)}$ which corresponds to an RSB function $\sigma_1(u)$. This
leads to an additional glass order parameter $\Delta _1(u)=
u\sigma_1 (u)-\int_{0}^{u}\sigma_1 (v)\,dv$ and 
the divergent $I(z)$ is replaced by a term in Eq.~(\ref{diag}) of 
the form
\begin {equation}
\sum_{{\bf q},k}\frac{1}{c(k)q^2+z}\int_{0}^{1}\frac{dv}{v^2}
\frac{\Delta_1(u)q^2/k_z^2}{c(k)q^2+z+\Delta(u)+\Delta_1(u)q^2/k_z^2}
\end{equation}
which converges at $k\rightarrow 0$. The general solution for both 
$\Delta(u)$ and $\Delta_1(u)$ involves a rather difficult set of two 
coupled differential equations. For $J=v=0$ the Bragg glass solution 
is \cite{Giamarchi,Korshunov} $\Delta_1(u)\sim u^2$ for $u<u_c$ and 
$\Delta_1(u)=\Delta_1(u_c)$ for $u_c<u<1$, where $u_c\sim 1/s_0$ is 
small. Thus for $t=T/T_{d}\gg u_c$ the structure of $\Delta_1(u)$ 
at small $u$ should not be affected by $\Delta(u)$ with its step at 
$u=2t$. The scale at which the $k$ divergence is cutoff is at 
$k<\Delta_1(u_c)/c(0)\approx 1/\ell _z$.

Consider then $I(z)$ with a lower cutoff $\pi /\ell _z$ on the 
$k$ integral. If $\ell _z<a$ it leads to a 
small correction $\sim O(d/\ell _z)$ to the main 
$1/a\lesssim k<\pi/d$ integration range. 
If $\ell _z >a$ then the $\pi/\ell _z <k\lesssim 1/a$ range in $I(z)$ can be 
neglected if 
\begin {equation}
\ell _z <(a^4/d^3)/[32ln^2(a/d)] .
\label{condition}\end{equation}
 Since $a\gg d$ we expect this to be valid; 
 in particular for BSCCO parameters (see 
below) it is always valid. Note also that a finite thickness of the 
sample can also serve as a cutoff replacing $\ell _z$.

$I(z)$ with $1/a\lesssim k<\pi /d$ integration yields 
$(s_0/8\pi ^2)I(z)=2s\log (\Delta _c/z)$ where the dimensionless 
disorder parameter is
$s=4\pi {\bar U}\lambda_{ab}^4/[\tau ^2a^2\log ^2 (a/d)]$ .
The renormalized Josephson coupling of Eq.~(\ref{extremumb}) is then
\begin {equation}
z/\Delta _c=e^{-1}[J^2/(8T^2 t\xi_0^2 d\Delta_c)]^{1/(1-2s)} .
\label{z}\end{equation}
Comparing Eqs.~(\ref {Delta},\ref{z}) shows that $\Delta_0$ vanishes at
$s=t$ (up to a nonuniversal $\sim 1/\ln v$ term) and formally there is a 
solution with $\Delta_0<0$ when $s<t$. However, the average 
distribution \cite{Mezard} of $|b({\bf q},k)|^2$ is $\sim \exp [-|b({\bf 
q},k)|^2/G_{\alpha,\alpha}({\bf q},k)]$ is acceptable only if 
$G_{\alpha,\alpha}({\bf q},k)>0$. This is a thermodynamic stability 
criterion and for our solution it reduces to $\Delta_0>0$. Thus the 
regime where both $z$, $\Delta_0$ are finite is limited to 
$s<\case{1}{2}$, 
$t<s$; we term this regime the Josephson Glass (JG) phase.
The glass parameter vanishes (continuously) at $t=s$ while the Josephson 
coupling vanishes (continuously) at $s=\case{1}{2}$.
 For $s>\case{1}{2}$ and $t<\case{1}{2}$ the solution is $z=0$
while $\Delta_0 \neq 0$ satisfies Eq.~(\ref{Delta}), i.e. it is a 
decoupled glass phase. Finally, for $\Delta_0=0$ a replica symmetric 
solution is valid at $s<t<1-s$ with
\begin {equation}
z/\Delta_c\approx (J/2T\xi_{0}^{2}d\Delta_c)^{1/(1-s-t)} .
\end{equation}
Thus $s+t=1$  for $s<\case{1}{2}$ defines a decoupling transition.

\begin{figure}[htb]
\begin{center}
\includegraphics[scale=0.7]{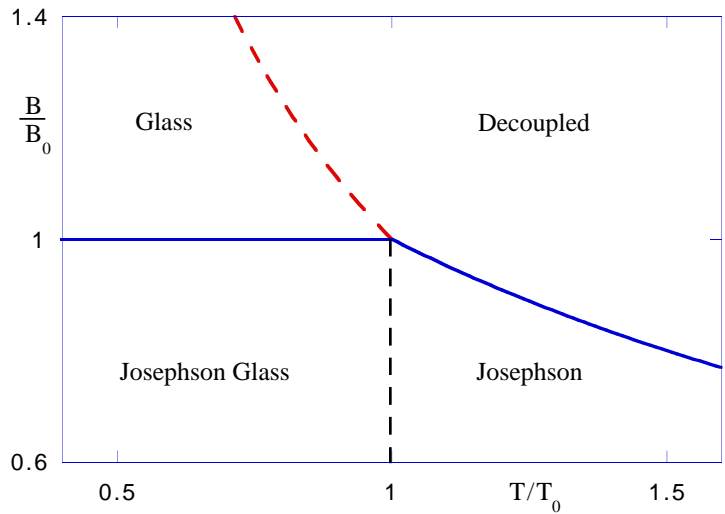}
\end{center}
\caption{Phase diagram. Full lines are decoupling lines where the Josephson 
coupling vanishes. Dashed lines are depinning lines where the 
Josephson glass parameter vanishes. $B_0$ is determined by the disorder 
strength while $T_0=\case{1}{2}T_d (a=a_0)$ (Eq.~(\ref{Td})) where 
$a_0^2=B_0/\phi_0$.}
\end{figure}

The phase diagram, Fig. 1, has four phases which all meet at a 
point defined as $B_0,T_0$. $B_0$ is determined by the disorder strength 
via $s=\case{1}{2}$ while $T_0=\case{1}{2}T_d (a=a_0)$ 
(Eq.~(\ref{Td})) where 
$a_0^2=B_0/\phi_0$. Since s increases with $B$ the 
$s=\case{1}{2}$ line 
defines a decoupling transition from a JG phase at low $B$ to a pinned 
glass phase (G) at high fields. It is remarkable that although the G phase 
has vanishing Josephson coupling ($z=0$) the Josephson induced disorder (the $v$ 
term in Eq.~(\ref{Hreplica})) is dominant in determining the glass 
nature of this phase. In fact, RG shows that J first 
increases (scaling from $\xi_0$ to $1/q_u$), generating the $v$ term, 
and only at scales beyond $1/q_u$ $J$ decreases to zero.

The JG phase at $B<B_0$ undergoes another transition at $t=s$, i.e. 
 at $T=T_0$ (up to $\ln B$ factors)
into a phase with finite Josephson coupling while the  
glass parameter $\Delta_0$ vanishes. This Josephson (J) phase has, 
however, the Bragg glass type disorder. The condition that $\ell _z$ has 
negligible effects in the JG or G phases implies that the pinning effect from 
the Josephson induced $\Delta_0$ is much stronger than that 
associated with the Bragg glass. Thus the JG-J transition is a 
depinning transition, from strong to weak pinning. The G phase 
also undergoes a depinning transition into a decoupled Bragg glass 
phase (D) at $T=B_0T_0/B$; the D phase is a Bragg glass phase maintained by
the interlayer e-m coupling.

The J phase undergoes a decoupling transition at 
 $B=2B_0T_0/(T+T_0)$, 
using $s\approx B/2B_0$. The J-D transition is continuous, 
at least for small J/T 
where RG and RSB are valid. The SCHA
 has been formally extended to higher $J/T$ 
and found to be of first order \cite{Daemen}. We recall now
that melting and related dislocations have been neglected. An upper 
bound on $T_m$ is the vortex transition \cite{Horovitz1} at $\tau /8$ 
where $p\overline {p}$ pairs are thermaly excited. Thus $T_0 
<\tau/8$ limits our description near $T_0$ to $a_0\lesssim 
\lambda_{ab}$.

We reconsider now the condition for the Bragg glass with its domain 
size $\ell _z$ to have 
a negligible effect on our glass phase. The coefficient of the 
$\cos [{\bf Q}\cdot ({\bf u}_{l}^{n,\alpha}-{\bf u}_{l}^{n,\beta})]$
term is related to $s_0$ and yields the Bragg glass length 
\cite{Giamarchi}
$\ell _z \approx 0.004\lambda_{ab}^2/[sd\log ^2(a/d)]$.
This domain size has non-negligible effects only if $\ell _z>a$ 
 and also if $\ell _z$ does not
satisfy Eq.~(\ref{condition}), i.e.
$2\cdot 10^3 a_0^2d/\lambda_{ab}^2< a <0.5d\lambda_{ab}/a_0$ with
$s\approx a_0^2/a^2$. This range exists if 
$a_0<0.06\lambda_{ab}$, i.e. $B_0>10^5 G$ for $\lambda_{ab} \approx 1700 
\AA$. Thus Bragg glass effects are negligible for BSCCO and YBCO, except 
near $B_0$ in the $B_0\approx 10 T$ YBCO sample \cite{Deligiannis}.
Note also that even if $a_0<0.06\lambda_{ab}$, Bragg glass
effects can be neglected if the sample thickness satisfies
  Eq.~(\ref{condition}).

We interpret the experimentally observed second peak phenomena 
\cite{Kes,Khaykovich1,Khaykovich2,Deligiannis,Giller} 
as the JG-G transition, i.e. a 
decoupling transition within the glass phase. While a decoupling 
scenario has been suggested as a crossover phenomena 
\cite{Kes,Koshelev1}, the present theory predicts a strict phase 
transition. The JG-G transition at $B=B_0$ is T independent up to 
$T_0$ and $B_0$ decreases with impurity strength; both features are 
consistent with experimental data \cite{Kes,Khaykovich1,Khaykovich2}.
The NCCO sample \cite{Giller} has similiar parameters to BSCCO 
samples implying a comparable $T_{0}$; however, the low value of 
$T_c\approx 23 K$ indicates 
that a multicritical point with $T_0<T_c$ is probably not realized. 

Recent data \cite{Fuchs,Kopelevich,Dewhurst} has shown an additional 
phase boundary in BSCCO, i.e. 
 a depinning transition line which crosses the critical point 
$B_0,\,T_0$. Our result for the depinning temperature, $T=T_0$  at $B<B_0$ 
being $B$ independent (up to $\sim \ln B$ terms),
is in accord with the data. At $B>B_0$ we expect the depinning line 
at $T=B_0T_0/B$, in qualitative agreement with a stronger $B$ 
dependence. \cite{Dewhurst}.

Neutron data \cite{Forgan} has shown a reentrant behavior in the 
$600-10^3 G$ range with positional correlations increasing with 
temperature. This is consistent with our decoupled phase which is 
weakly pinned, leading to enhanced positional correlations. The 
reentrant behavior seems to extend to $B<B_0$ so that our J-D line 
may be the first order line, at least near $B_0$; at lower fields 
this decoupling line probably joins the melting line.

In conclusion, we have found a phase diagram which is remarkably 
close to the experimental one
\cite{Kes,Khaykovich1,Khaykovich2,Deligiannis,Fuchs,Dewhurst}, having a multicritical 
point where four phases meet. Our theory provides a fundamental 
interpretation of both the second peak transition and the recently 
discovered depinning transition.

\vspace{4mm}
{\bf Acknowledgments}: We thank E. Zeldov, D. T. Fuchs for 
most valuable and stimulating discussions.
This research was supported by THE ISRAEL SCIENCE FOUNDATION founded 
by the Israel Academy of Sciences and Humanities.

\end{document}